\documentclass[12pt]{article}
\usepackage{amsmath, euscript, amssymb, amsfonts}

\newcommand{\oR}{{\mathbb R}}
\newcommand{\oV}{{\mathbb V}}
\newcommand{\oC}{{\mathbb C}}

\newcommand{\oK}{{\mathbb K}}
 
\newcommand{\oT}{{\mathbb T}}

\newcommand{\oJ}{{\mathbb J}}
\newcommand{\oU}{{\mathbb U}}

\newcommand{\supp}{\mathop{\rm supp}\nolimits}
\newcommand{\Spec}{\mathop{\rm Spec}\nolimits}

\newcommand{\R}{\mathop{\rm Re}\nolimits}
\newcommand{\I}{\mathop{\rm Im}\nolimits}

\begin{document}

\hfill  FIAN/TD/7-06

\bigskip

\hfill hep-th/0605249

\vspace{2cm}

\begin{center}

{\Large\bf Axiomatic formulations  of  nonlocal and

\medskip noncommutative field theories}

\vspace{1cm}

{\bf M.~A.~Soloviev}\footnote{E-mail: soloviev@lpi.ru}

\vspace{0.5cm}

{\sl P.~N.~Lebedev Physical Institute

Russian Academy of Sciences

Leninsky Prospect 53, Moscow 119991, Russia}

\vspace{1cm}

{\bf Abstract}

\end{center}

 We analyze   functional analytic  aspects of
axiomatic formulations of nonlocal and noncommutative quantum
field theories. In particular, we completely clarify  the relation
between the asymptotic commutativity condition, which ensures the
CPT symmetry and the standard spin-statistics relation for
nonlocal fields, and the regularity properties of the retarded
Green's functions in momentum space that are required for
constructing  a scattering theory and  deriving  reduction
formulas. This result is based on a relevant
Paley-Wiener-Schwartz-type theorem for analytic functionals. We
also discuss  the possibility of using analytic test functions to
 extend  the Wightman axioms to noncommutative field theory,
where the causal structure with the light cone is replaced by that
with the light wedge. We explain some essential peculiarities of
deriving  the CPT and spin-statistics theorems in this enlarged
framework.


\vskip 2em

\noindent
{\bf Keywords:} nonlocal quantum fields, causality,
noncommutative field theory,  Wightman functions, analytic
functionals, Paley-Wiener-Schwartz  theorem

\vskip 2em

 \section{\large  Introduction}
This paper is concerned with some questions that arise in
extending the Wightman axioms to nonlocal and noncommutative
quantum field theories (QFTs) and are related to using enlarged
classes of generalized functions suggested for this purpose
instead of the tempered distributions used in the conventional
formalism~\cite{J}-\cite{BLOT}.  The generalized functions defined
on the space $S^0$ of analytic test functions were perhaps
proposed most often. This space is simply the Fourier transform of
the Schwartz space $\mathcal D$ of infinitely differentiable
functions of compact support and can serve as a functional domain
 for fields with an arbitrary high-energy behavior. The
first  question discussed below has a long history. In~\cite{St2},
Steinmann proposed  replacing the local commutativity axiom, which
loses its meaning for  fields defined on  $S^0$, by some
regularity conditions  on the retarded Green's functions in
momentum space that ensure the existence of an $S$-matrix and
provide a basis for  developing  a scattering theory. More
recently, a generalization of microcausality  in terms of
coordinate space was found. This generalization,  called
asymptotic commutativity, ensures the standard spin-statistics
relation and the CPT invariance of nonlocal QFT (see~\cite{1} and
the references therein). The asymptotic commutativity is stated as
the continuity property of the commutators of  observable fields
with respect to the topology of a space related to  $S^0$ and
associated with the closed light cone. In~\cite{S}, a
Paley-Wiener-Schwartz-type theorem was proved for the generalized
functions defined on $S^0$, which allows  clarifying  the relation
between  asymptotic commutativity and Steinmann's conditions. But
doing this, as shown below,  requires  further extending  the
theorem to multilinear forms.

The same mathematical technique proves  useful for analyzing
recent axiomatic formulations of noncommutative field theory. As
noted in~\cite{*}, such an axiomatization is somewhat premature
because of the shortage of well-studied instructive examples.
Actually, there is no general agreement regarding the situation
with the Poincar\'e symmetry, causality and unitarity in
noncommutative field theory, and  this name is often used for
theories with quite different physical contents. Here we discuss
attempts to extend the Wightman axioms to field theories obtained
from the Seiberg-Witten limit~\cite{SeiW} of string theories; a
feature of these field theories is the replacement of the causal
structure with the light cone by that with the light wedge. Such a
modification of the microcausality condition was proposed, for
example, in~\cite{Alv} and was examined in more detail
in~\cite{CFI},~\cite{HRR}. An analysis of this formulation is also
 interesting from the methodical standpoint. It shows that there
is no need to use the whole  Bargmann-Hall-Wightman theorem to
derive the spin-statistics relation and the CPT invariance because
it suffices to use its simplest version for the two-dimensional
space-time, where the complex Lorentz group coincides with the
group $\oC^*=\oC\setminus \{0\}$ of complex numbers.
 It was emphasized in~\cite{Alv},~\cite{HRR} that
the framework of tempered distributions is hardly suitable for the
theory developed by the authors in view of  the singularities
related to UV/IR mixing and because of the exponential growth of
the momentum-space correlators  along the noncommutative
directions. In~\cite{FP},  generalized functions of the class
$S^{\prime\, 0}$ were proposed instead. In the second part of the
present paper, we explain some subtleties in proving  the CPT and
spin-statistics theorems in such an enlarged framework and show
that the most straightforward derivation of these theorems is by
extending the  uniqueness theorem obtained in~\cite{2} for the
distributions supported in a properly convex cone to the case of a
wedge-shaped support.

Section 2 contains the necessary information about the functional
spaces used. In the same section, we introduce the key notions of
carrier cones of a functional belonging to $S^{\prime\, 0}(\oR^d)$
and of  a multilinear form defined on $S^0(\oR^d)\times\dots\times
S^0(\oR^d)$. In Sec.~3, we state Steinmann's regularity conditions
 and briefly describe the characteristic properties of the retarded
products of nonlocal quantum fields. In Sec.~4, we obtain a
Paley-Wiener-Schwartz-type theorem for multilinear forms and use
this result  to establish the correspondence between  Steinmann's
conditions and  asymptotic commutativity. In Sec.~5, we  extend
the uniqueness theorem obtained in~\cite{2} to distributions
supported in a properly convex wedge and show that this allows
deriving analogues of the CPT and spin-statistics theorems for
theories with space-space noncommutativity in exactly the same
manner as for nonlocal QFT in~\cite{1}. Section~6 contains
concluding remarks.

\section{\large Preliminaries}

We recall that  $S^0$ is the space of all entire analytic
functions satisfying the inequalities
\begin{equation}
|f(x+iy)|\le C_N(1+|x|)^{-N}\,e^{B|y|},\quad N=0,1,2,\dots,
 \notag
   \end{equation}
 where $C_N$ and $B$ are constants depending on $f$~\cite{GS}. To develop a
 field theory with test functions of this kind, certain spaces
 related to $S^0$ and associated with cones in $\oR^d$ should be used. If
 $U\subset\oR^d$ is an open cone, then the space $S^0(U)$ is defined
 as the union of the countably normed spaces
$S^{0,b}(U)$, $b>0$, consisting of entire functions with finite
norms of the form
\begin{equation}
     \|f\|_{U,B,N}=  \sup_{z\in
\oC^d}|f(z)|\prod_{j=1}^d(1+|x_j|)^N e^{- Bd(x,U) -B|y|},\quad
B>b,\,\,N=0,1,2,\dots,
 \label{1}
    \end{equation}
    where $z=x+iy$ and $d(x,U)=\inf_{\xi\in U}|x-\xi|$ is the
    distance of $x$ from  $U$. Clearly, $S^0(U)$ is continuously
    embedded in    $S^0(U')$ if $U\supset U'$.
     Hereafter, we take the norm in $\oR^d$ to be $|x|=\sum_j|x_j|$.
     Norm~\eqref{1} then has the nice property  of multiplicativity, which
     is used below in treating tensor products. Namely, if
 $f_1\in S^0(U_1)$ and $f_2\in S^0(U_2)$, then
\begin{equation}
     \|f_1\otimes f_2\|_{U_1\times U_2,B,N}=
      \|f_1\|_{U_1,B,N}\cdot\|f_2\|_{U_2,B,N}.
 \label{m}
    \end{equation}
We let  $S^{\prime\, 0}$  denote the space of continuous linear
functionals on $S^0$ and call a closed cone $K\subset \oR^d$  a
carrier of $v\in S^{\prime\, 0}(\oR^d)$ if the functional $v$
allows a continuous extension to each space $S^0(U)$, where
$U\supset K\setminus \{0\}$ (the last inclusion is also written as
$U\Supset K$). This amounts to saying that $v$ has a continuous
extension to the inductive limit
  \begin{equation}
S^0(K)=\varinjlim_{U\Supset K}S^0(U).
    \label{2}
   \end{equation}
Let    $w$ be a separately continuous  multilinear form on
 $\underbrace{S^0(\oR^{d})\times\dots \times S^0(\oR^{d})}_n$ and let
 $K=K_1\times\dots\times K_n$, where $K_j$ are closed cones in
   $\oR^{d}$. It is natural to say that $w$ is carried by $K$
    if every $K_j$  is a carrier of all linear functionals
   on $S^0(\oR^{d})$   defined by
     $w(f_1,\dots,f_n)$, where $f_i$ with
    $i\neq j$ are held fixed. By Schwartz's kernel theorem, for each form
    $w$, there exists a unique linear functional
    $v\in S^{\prime\, 0}(\oR^{dn})$   such that
    $$
    w(f_1,\dots,f_n)=v(f_1\otimes\dots\otimes f_n).
    $$
In the field theory context, this means that the vacuum
expectation values of products of fields are identified with
certain generalized functions, just as in the usual
formalism~\cite{J}-\cite{BLOT}. In considering the questions
stated in the introduction, we use the following theorems.

\medskip
 {\bf Theorem 1.} {\it The space  $S^0(\oR^d)$ is dense in
 every space $S^0(U)$, where $U$ is an open cone in
$\oR^d$.}

\medskip
 {\bf Theorem 2.} {\it Every element in $S^{\prime\, 0}(\oR^d)$
has a unique minimal carrier cone.}

\medskip
{\bf Theorem 3}. {\it If a functional $v\in S^{\prime\, 0}(\oR^d)$
is carried by a properly convex cone $K$, then it has the Laplace
transform $\mathbf u(\zeta)=(v, e^{iz\zeta})$, which is an
analytic function on the tube $T^V=\oR^{d}+iV$, where $V$ is the
interior of the dual cone $K^*=\{\eta\colon x\eta\ge 0, \forall
x\in K\}$. This function satisfies the condition
 \begin{equation}
  |\mathbf u(\zeta)|\leq
 C_{R,V'}\, |\I \zeta|^{-N_{R, V'}},\qquad
  \I \zeta\in V',\,\, |\zeta|\leq R,
 \label{3}
 \end{equation}
for every $R >0$ and every $V'\Subset V$. If ${\rm
Im}\,\zeta\rightarrow 0$ in a fixed  $V'$, then ${\mathbf
u}(\zeta)$ tends to the Fourier transform of $v$ in the topology
of $\mathcal D'$. Conversely, every function that is analytic on
 $T^V$, where $V$ is an open cone in $\oR^d$, and that satisfies~\eqref{3}
 is the Fourier-Laplace transform of an element in $S^{\prime\, 0}(V^*)$.}

\medskip
{\bf Theorem 4}. {\it A separately continuous multilinear form on
$\underbrace{S^0(\oR^{d})\times\dots \times S^0(\oR^{d})}_n$ is
carried by a cone $K_1\times\dots\times K_n$ if and only if its
associated generalized function on $\oR^{dn}$ has a continuous
extension to the space}
\begin{equation}
S^0(K_1,\dots, K_n)=\varinjlim_{U_1\Supset K_1,\dots, U_n\Supset
K_n }S^0(U_1\times\dots\times U_n).
    \label{4}
   \end{equation}

\medskip
These theorems are similar to those previously established for
another functional class $S^{\prime\, 0}_\alpha$, but  proving
them is somewhat more laborious because the topological structure
of $S^0(U)$ is more complicated than that of $S^0_\alpha(U)$.
 Theorem 1 is derived in~\cite{2} from a density
theorem for $S^0_\alpha(U)$. Theorems 2 and 3 are established
in~\cite{S}. Definition~\eqref{4} is an analogue of the definition
given in~\cite{Sm} for $S^0_\alpha(K_1,\dots,K_n)$. Theorem~4 is
derivable in the same manner as Theorem 3 in~\cite{S4} but using
Theorem 1 in~\cite{S5} instead of Lemma 5 in~\cite{S4}. For the
applications under consideration, it is  essential that the space
$S^0(K_1,\dots, K_n)$ does not  coincide with
$S^0(K_1\times\dots\times K_n)$ in general. The latter  is a
subspace of $S^0(K_1,\dots, K_n)$  but  a proper subspace unless
$K_j=\{0\}$ and $K_j=\oR^{d}$ for all $j$
(see~\cite{Sm},~\cite{S4}). If a functional $v\in
S^{\prime\,0}(K_1\times\dots\times K_n)$ extends to
$S^0(K_1,\dots, K_n)$, we say that $v$ is strongly carried by the
multiple cone $K_1\times\dots\times K_n$.

 \section{\large Steinmann's  condition}

We consider the simplest case of a  neutral nonlocal scalar field
$\phi$. This field is assumed to be an operator-valued generalized
function defined on $S^0(\oR^4)$ instead of the space $S(\oR^4)$
of rapidly decreasing smooth functions, which is commonly used in
local QFT~\cite{J}-\cite{BLOT}. We assume that $\phi$ satisfies
all the Wightman axioms except  the microcausality condition which
cannot  be   formulated in terms of such test functions. Let $D$
denote a common dense domain of the operators $\phi(f)$, $f\in
S^0(\oR^4)$, in the Hilbert space of states  and let
$U(\Lambda,a)$ be a unitary representation of the Poincar\'e group
in this space. As in~\cite{St2}, we assume that there exist
 retarded products $R(x;x_1,\dots,x_n)$, $n=1,2,\ldots$, of
fields with the following characteristic properties:
 \begin{itemize}

\item[1.]
 the operators
 $$
 R(f)=\int\!
R(x;x_1,\dots,x_n)\,f(x,x_1,\dots,x_n)\,{\rm d}x{\rm d}x_1\dots
{\rm d}x_n,\quad f\in S^0(\oR^{4(n+1)}),
$$
are defined on the domain $D$ and map it into itself,

\item[2.)]
 $R(f)$ are Hermitian for real  $f$,
 \item[3.]
 $R(x)=\phi(x)$,

\item[4.]
 the $R$ products are symmetric in the variables $x_1,\dots,x_n$,

\item[5.] the equality
\begin{multline*}
 R(x;y,x_1,\dots,x_n) - R(y;x,x_1,\dots,x_n)=\\
=-i\sum[R(x;x_{i_1},\dots,x_{i_\alpha}),R(y;x_{i_{\alpha+1}},\dots,x_{i_n})]
\end{multline*}
 holds,  where the sum ranges  all partitions of
$\{x_1,\dots,x_n\}$ into two subsets,

 \item[6.]    $R(\Lambda x+a;\Lambda x_1+a,\dots, \Lambda x_n+a)=U(\Lambda,
  a)R(x;x_1,\dots,x_n)U(\Lambda,a)^{-1}$.
  \end{itemize}

\noindent
 In local field theory, a leading role is played by the
causality condition
 \begin{itemize}
  \item[7.]  $\supp R(x;x_1,\dots,x_n)\subset \oK_n$,  where
    \begin{equation}
  \oK_n= \oR^4\times \bar
\oV_-\times\dots\times \bar \oV_-=\{(x,x_1,\dots,x_n)\colon
(x_j-x)\in \bar \oV_-,\, j=1,\dots,n\}
    \label{2.1}
  \end{equation}
  and $\bar \oV_-=\{\xi\colon\,\xi^2\ge0,\xi^0\le 0\}$ is the
  closed backward  light cone.
\end{itemize}

   Steinmann proposed replacing  support property 7
    with a regularity condition  for the  retarded products in $p$-space.
   The Fourier  transform of  $\langle\Phi,
  R(x;x_1,\dots,x_n)\Psi\rangle$, where $\Phi,\Psi\in D$, can be
  regarded as a distribution in the
   variables  $p_1,\dots,p_n$,  and $P=p+p_1+\dots,+p_n$. Let
  $\chi(p_1,\dots, p_n)$ denote the distribution resulting from
  the integration with a test function in $P$. Then the regularity
  condition  is written as follows.\footnote{In~\cite{St2}, this
   condition was stated for  generalized retarded products,
   but we  restrict our consideration to the simplest case of
    ordinary retarded products.}

\medskip
  {\bf Condition $\mathcal R$.} {\it The function
    $\chi(p_1,\dots, p_n)$ is analytic on the tubular domain
\begin{equation}
\oT_-^n=\{(p_1,\dots,p_n)\colon \I p_j\in \oV_-,\,
  j=1,\dots,n \}.
 \label{2.2}
  \end{equation}
  For each compact set $Q\subset \oR^{4n}$, for every
  $\gamma>1$, and for every $R>0$, there exist a positive
   constant $C$ and nonnegative integers $N_j$ such that
 \begin{equation}
  |\chi(p_1,\dots, p_n)|\le C\prod_{j=1}^n |\I p_{j0}|^{-N_j}
  \label{2.3}
  \end{equation}
  for  $(p_1,\dots,p_n)\in \oT_-^n$, $(\R p_1,\dots,\R p_n)\in Q$
   and $\gamma |\I \mathbf p_j|\le |\I p_{j0}|\le
  R$, $j=1,\dots,n$  $($here $p_{j0}$ is the $0$-component
  of the four-vector $p_j$
  and $\mathbf p_j$ is its spatial part$)$}.

\medskip
By this condition,  the analyticity properties of the retarded
products of nonlocal fields in $p$-space are
  identical to those of the retarded products in the conventional
  local formalism~\cite{BLOT}. Restriction~\eqref{2.3}
ensures that the analytic function $\chi$ has a boundary value
belonging to the space $\mathcal D'$ of Schwartz distributions.
But in contrast to the local theory, no restrictions are imposed
on the behavior of $\chi$ for large $p_j$, real or imaginary.

  The problem of clarifying the meaning  of Condition  $\mathcal R$
   in coordinate space was not posed in~\cite{St2}. The author of
   that paper  emphasized that this condition,
   in his opinion,  is too complicated
  and  too far from any direct physical interpretation to be accepted
  as a basic postulate. He suggested regarding it simply as
   a convenient technical property sufficient
  for  developing a consistent scattering formalism for
  nonlocal fields, including  the proof that
   asymptotic states and  the $S$-matrix exist.  The main result of~\cite{St2} is
   that under Condition   $\mathcal R$ the
   Lehmann-Symansik-Zimmermann reduction formulas for the $S$-matrix
   elements   hold in exactly the same form as in local QFT.
   We  now show that this condition when translated into
   $x$-space is a natural   generalization of  support property~7
   of the retarded products of local fields and can therefore be
   considered as  an appropriate candidate for the causality condition in
   nonlocal field theory. We also note that an analogue of the time-ordered
$T$-product can be constructed from  the $R$-products using the
usual recursive relations~\cite{BLOT} and Condition $\mathcal R$
implies that there  exists a single analytic function whose
boundary values from different domains  are the  Fourier
transforms of the retarded and causal  Green's functions
$\langle\Psi_0, R(x;x_1,\dots,x_n)\Psi_0\rangle$ and
$\langle\Psi_0, T(x;x_1,\dots,x_n)\Psi_0\rangle$, where $\Psi_0$
is the vacuum state.

\section{\large Relation to asymptotic commutativity}

Theorem 4 shows that the space of  separately continuous
multilinear forms defined on  $S^0(\oR^{d})\times\dots \times
S^0(\oR^{d})$ and  carried by a multiple cone
$K_1\times\dots\times K_n$ is identified with the dual of
space~\eqref{4}. Let $V_1,\dots, V_n$ be open cones in
    $\oR^d$ and let $\mathcal A_0(V_1,\dots, V_n)$ denote the
    space of all functions analytic on the tube
      $T^V=\oR^{dn}+iV$, where  $V=V_1\times\dots\times V_n$,
and satisfying
 \begin{equation}
  |\mathbf u(\zeta)|\leq
 C_{R,V'_1,\dots,V'_n}\prod_{j=1}^n |\I \zeta_j|^{-N_{R, V'_j}},\qquad
   \I \zeta_j\in V'_j,\,\, |\zeta_j|\leq R,\,\, j=1,\dots,n,
 \label{6.1}
 \end{equation}
for each $R>0$ and for every $V'_j\Subset V_j$. Clearly, $\mathcal
A_0(V_1,\dots, V_n)$ is an algebra under   pointwise
multiplication.

\medskip
 {\bf Theorem 5.} {\it  The Laplace transformation
  $\mathcal L:\,v\to (v, e^{iz\zeta})$ is an isomorphism of the
  space  $S^{\prime\,0}(V_1^*,\dots, V_n^*)$ onto the algebra
   $\mathcal A_0(V_1,\dots, V_n)$. If $\I \zeta\to 0$ inside a
   fixed cone $V'_1\times\dots\times V'_n$, where
$V'_j\Subset V_j$, then the function $(\mathcal Lv)(\zeta)$ tends
to the Fourier transform of  $v$ in the topology of $\mathcal
D'(\oR^{dn})$.}

\medskip
{\bf Proof.} Because $S^{\prime\,0}(V_1^*,\dots, V_n^*)\subset
S^{\prime\,0}(V^*)$, we can apply Theorem~3, whose statement
coincides with that of Theorem~5 for $n=1$ and which, in
particular, shows that every functional in $S^{\prime\,0}(V^*)$
has a Laplace transform analytic on $T^V$. The bound
in~\eqref{6.1} is stronger then the bound in~\eqref{3}, which
holds for an arbitrary element in $S^{\prime\,0}(V^*)$, but can be
obtained the same way starting from the  estimate
$$
|\mathcal Lv(\zeta)| =|(v,e^{iz\zeta})|\le
\|v\|_{U,B,N}\|e^{iz\zeta}\|_{U,B,N},
 $$
where $U=U_1\times\dots\times U_n$, with $U_j$ being any open
cones such that $V_j^*\Subset U_j$, $B$ can be taken arbitrarily
large, and $N$ generally depends  on $B$ and $U$. Because
of~\eqref{m}, we have
$$
\|e^{iz\zeta}\|_{U,B,N}=\prod_j\|e^{iz_j\zeta_j}\|_{U_j,B,N},
 $$
 where each factor can be estimated  exactly the same way
 as the norm of an exponential
 in~\cite{S}, which yields~\eqref{6.1}.

  We now let $u$ be the boundary value of a function $\mathbf u$
with property~\eqref{6.1}, which exists in $\mathcal D'(\oR^{dn})$
by Theorem 3.1.15 in~\cite{H1}. Theorem 3 shows that the cone
$V_1^*\times\dots\times V_n^*$ is a carrier of the multilinear
form defined on
 $S^0(\oR^{d})\times\dots \times S^0(\oR^{d})$ by the inverse
 Fourier transform of the distribution $u$. Applying Theorem 4
 finishes the proof.

By rewriting the regularity condition for the Green's functions of
the nonlocal field theory in terms of Theorem 5, we obtain the
following result.

\medskip
 {\bf  Theorem 6.} {\it
Condition $\mathcal R$ is equivalent to the requirement that in
$x$-space, the functionals defined on $S^0(\oR^{4(n+1)})$ by the
matrix elements $\langle\Phi, R(x;x_1,\dots,x_n)\Psi\rangle$,
$\Phi,\Psi\in D$, are strongly carried by  cone~\eqref{2.1}.}

\medskip
More specifically, Condition $\mathcal R$ means that all these
functionals allow a continuous extension to any space
$S^0(U_\gamma)$,
 where $$
 U_\gamma=\{(x,x_1,\dots,x_n)\colon
|\mathbf{x-x}_j|<\gamma(x^0-x_j^0),\, j=1,\dots,n\},\quad
\gamma>1.
$$
In particular, when coupled with  property
 5  in Sec.~3, this condition implies that the matrix
 elements of the commutator
$$
\langle\Phi, [\phi(x),\phi(y)]\Psi\rangle=\langle
\Phi,(R(x,y)-R(y,x))\Psi\rangle,\qquad \Phi,\Psi\in D,
$$
are strongly carried by the cone  $\oR^4\times \bar \oV=\{(x,y)\in
\oR^8\colon (x-y)^2\geq 0\}$. Therefore the theory satisfies the
asymptotic commutativity condition~\cite{1}, which ensures the CPT
invariance and the standard spin-statistics relation.

\section{\large An axiomatic formulation of noncommutative QFT}

We consider a theory with space-space noncommutativity, namely,
with the commutation relation
\begin{equation}
  [\hat x^\mu,\hat x^\nu]=i\theta^{\mu\nu},
 \label{7.1}
 \end{equation}
 where $\theta^{23}=-\theta^{32}$ is a constant noncommutativity
 parameter and all other matrix entries $\theta^{\mu\nu}$ are zero.
  Clearly, in the Lorentz group, $O(1,1)\times SO(2)$
  is the   largest subgroup
 leaving  relation~\eqref{7.1} invariant. Its identity component is the
 product of the group $SO_0(1,1)$ of boosts in the plane $(x^0,x^1)$
   and the group $SO(2)$ of rotations in the
  plane $(x^2,x^3)$. The main idea in~\cite{Alv} is to match
  the Wightman axioms with this residual symmetry. From
  such a standpoint, letting $\mathcal T_4$ denote the group of
  space-time translations, we assume the existence of a continuous unitary
  representation $U(\Lambda,a)$ of the connected group
  \begin{equation}
[SO_0(1,1)\times SO(2)]\rtimes \mathcal T_4,
 \label{7.2}
 \end{equation}
in the Hilbert space of states and suppose that there is a unique
vacuum vector $\Psi_0$, which is invariant under $U(I,a)=e^{ia^\mu
P_\mu}$. The modified spectral condition proposed in~\cite{Alv}
implies that the spectrum of the energy-momentum operator $P$ lies
in the forward light wedge, i.e.,
\begin{equation}
\Spec P\subset\bar \oV_{c+}\times \oR^2=\{p\in \oR^4\colon\,
p_c^2= p_0^2-p_1^2\ge 0, p_0\ge 0\}.
 \label{7.3}
\end{equation}
Hereafter, we use the subscript $c$ to denote the restriction of
variables to the commutative subspace.
 Analogously, the microcausality axiom is replaced by the condition
 that the field commutators (or anticommutators for unobservable fields)
  $[\phi_\iota(x),\phi_{\iota'}(x')]_{\stackrel{-}{(+)}}$
  vanish outside the closed wedge
 \begin{equation}
 \bar \oV_c\times\oR^6=\{(x,x') \colon \,
 (x-x')_c^2\geq 0\}.
 \label{7.4}
\end{equation}
We  restrict our consideration to the case of a complex scalar
field $\phi$ with the transformation law
\begin{equation}
U(\Lambda,a)\phi(f)U^{-1}(\Lambda,a)=\phi(f_{(\Lambda,a)}),
\label{7.5}
\end{equation}
where $f_{(\Lambda,a)}(x)=f(\Lambda^{-1}(x-a)$, and we comment on
the possibility of  generalizing this axiomatic framework  using
analytic test functions in $S^0$, as  proposed in~\cite{FP}.

We first point out a consequence of  conditions~\eqref{7.3},
\eqref{7.5} that holds independently  of a causality formulation.
Because of the translation invariance, the $n$-point vacuum
expectation value
$$
\mathcal W(x_1,\dots,x_n)=\langle\Psi_0,\, \phi^{(*)}(x_1)\dots
\phi^{(*)}(x_n)\Psi_0\rangle,
$$
where $\phi^{(*)}$  is either $\phi$ or the Hermitian adjoint
field $\phi^*$, can be identified with a generalized function
$W\in S^{\prime\,0}(\oR^{4(n-1)})$ in the relative coordinates
$\xi_j=x_j-x_{j+1}$, $j=1,\dots,n-1$. Lemma~4 in~\cite{1} shows
that the Wightman function  $W$ is carried by a cone $K\subset
\oR^{4(n-1)}$  if and only if $\mathcal W$ is strongly carried by
its inverse image $K\times \oR^4$ in $\oR^{4n}$ and the membership
relation $W\in S^{\prime\, 0}(K_1,K_2)$ amounts to $\mathcal W\in
S^{\prime\, 0}(K_1,K_2,\oR^4)$.
 We use the notation
\begin{equation}
\oJ^c_{n-1}=\oV_{c,R}^{n-1}\cup\oV_{c,L}^{n-1},
 \label{7.6}
\end{equation}
where  $\oV_{c,R}=\{(\xi^0,\xi^1)\colon\,
 (\xi^0)^2-(\xi^1)^2<0,\xi^1>0\}$ and $\oV_{c,L}=-\oV_{c,R}$.
 According to~\cite{J},  two-component open cone~\eqref{7.6}
 consists of the real points of analyticity of the Wightman functions of
  local field theory in a space-time of two dimensions.

\medskip
{\bf  Theorem 7.} {\it Let $\phi(x)$ be a scalar field defined as
an operator-valued generalized function on the space $S^0(\oR^4)$
and satisfying conditions~\eqref{7.3} and \eqref{7.5}. Then the
difference
\begin{equation}
 W(\xi_1,\dots,\xi_{n-1})- W(-\xi_1,\dots,-\xi_{n-1}).
  \label{7.7}
\end{equation}
is strongly carried by the wedge $\complement \oJ^c_{n-1}\times
\oR^{2(n-1)}$. In particular, the vacuum expectation value of the
commutator
\begin{equation}
 \langle\Psi_0,\, [\phi(x),\phi(x')]_-\Psi_0\rangle.
\end{equation}
is strongly carried by  wedge~\eqref{7.4}.}

\medskip
{\bf Proof.} The Wightman functions  are invariant under the joint
inversion of $\xi^2$ and $\xi^3$, which is implemented by the
$SO(2)$ rotation through $180^0$. Therefore, it suffices to prove
that the specified wedge is a strong carrier of the difference
\begin{equation}
W(\xi)- W(I_c\xi),
  \label{7.8}
\end{equation}
where $I_c$ is the inversion of the commutative coordinates
$\xi^0$, $\xi^1$ and the notation $\xi=(\xi_1,\dots,\xi_{n-1})$ is
used for brevity. By Theorem 4, this is equivalent to the
statement that the cone $\complement \oJ^c_{n-1}$ is a carrier of
all functionals obtained from~\eqref{7.8} by averaging with test
functions depending only on $\xi^2$ and $\xi^3$. Let $W_g$ be the
result of averaging $W$ with such a function $g\in
S^0(\oR^{2(n-1)})$. By the hypothesis of the theorem, the
functional $W_g\in S^{\prime\,0}(\oR^{2(n-1)})$ is invariant under
the proper orthochronous Lorentz group $SO_0(1,1)$ in $1+1$
dimensions, and its Fourier transform has support in the properly
convex cone $\bar \oV^{(n-1)}_{c+}$. If $W$ is assumed to be a
tempered distributions as in~\cite{Alv}, then this support
property implies that $W_g$ is the boundary value of a function
$\mathbf W_g(\zeta)$ analytic in the tube
$\oT_{c-}^{n-1}=\{\zeta=\xi_c+i\eta\colon\,\eta\in
\oV_{c-}^{n-1}\}$. By the Bargmann-Hall-Wightman
theorem~\cite{J}--\cite{BLOT}, the function $\mathbf W_g$ allows
an analytic continuation to an extended tube, and after
continuation, it is invariant under the proper complex Lorentz
group. In the two-dimensional case, this group coincides with the
multiplicative group $\oC^*=\oC\setminus\{0\}$ of complex numbers,
and if we use the light-cone coordinates $\zeta^\pm=(\zeta^0\pm
\zeta^1)/\sqrt{2}$, then we can write its action as
\begin{equation}
\zeta^\pm\to z^{\pm1}\zeta^\pm,\qquad z\in \oC^*.
  \label{7.9}
\end{equation}
 In particular, the complex group contains the inversion $I_c$, although
 this transformation does not belong to the identity component of
$SO(1,1)$.   Cone~\eqref{7.6} is obviously contained in the
extended tube\footnote{We note that the description~\cite{J} of
the real points of analyticity for any dimension has been obtained
just by reducing to this simple two-dimensional case.} because
$\oV_{c,R}^{n-1}= \{\xi_c\colon\, \xi^+>0, \xi^-<0\}$ and the
transformation~\eqref{7.9} with $z=-i$ carries $\oV_{c,R}^{n-1}$
into $i\oV_{c-}^{n-1}$. For $\oV_{c,L}^{n-1}$, the same role is
played by $z=i$. Because  $g$ is arbitrary,  this  immediately
implies that in the noncommutative theory~\cite{Alv}, which
assumes tempered distribution fields, the difference~\eqref{7.8}
is identically zero in the wedge $\oJ^c_{n-1}\times \oR^{2(n-1)}$.

The ultraviolet behavior of $W_g\in S^{\prime 0}$ can be
regularized by multiplying its Fourier transform by an
$SO(1,1)$-invariant function $\omega(P^2_c/\Lambda)$, where
$\omega\in C^\infty_0(\oR)$ and $P_c=\sum_j p_{c,j}$. Theorem 8
in~\cite{S3} shows that this yields a tempered distribution. For
brevity, let $G$ denote  the difference~\eqref{7.8} averaged with
$g$, and let $G_\Lambda$ be the result of its regularization.
According to the aforesaid, $\supp G_\Lambda$ is contained in the
complement of the Jost cone $\oJ^c_{n-1}$. Therefore, for any
$\Lambda$, this distribution has a continuous extension $\hat
G_\Lambda$ to the space $S^0(U)$, where $U\Supset \complement
\oJ^c_{n-1}$, and even to $S^0(\oU)$, where $\oU$ is the interior
of the closed cone $\complement \oJ^c_{n-1}$. Such an extension
can be defined by $(\hat G_\Lambda,f)=(G_\Lambda,\chi f)$, where
$\chi$ is an infinitely differentiable function that equals
 unity in an $\epsilon$-neighborhood of $\oU$  and vanishes outside the
 $2\epsilon$-neighborhood. Clearly,  multiplication by
  $\chi$ maps $S^0(\oU)$ continuously into the Schwartz space
 $S(\oR^{2(n-1)})$.  We can now apply the arguments used
 to prove the CPT theorem for nonlocal QFT in \cite{2}.
Namely, we choose $\omega$ such that $\omega(t)=1$ for $|t|\le 1$
and define a functional $\hat G$ on $S^0(\oU)$ by
\begin{equation}
(\hat G, f)= (\hat G_\Lambda, f),\quad \mbox{where}\quad
\Lambda=4\sqrt{2n^3}eb\quad \mbox{if}\quad f\in S^{0,b}(\oU).
  \label{7.10}
\end{equation}
Theorem 1 ensures that $\hat G$ is well defined. According to its
detailed formulation given in~\cite{2}, the space
$S^{0,b'}(\oR^d)$ is dense in $S^{0,b}(U)$ in the topology of
$S^{0,b'}(U)$ for each open cone $U\subset\oR^d$ and for every
pair of positive numbers $b$ and $b'$ such that $b'>2deb$. Hence,
there exists a sequence $f_\nu\in S^{0,4neb}(\oR^{2(n-1)})$ such
that $f_\nu\to f$ in $S^{0,4neb}(\oU)$. By the
Paley-Wiener-Schwartz theorem (Theorem 7.3.1 in~\cite{H1}), the
Fourier transforms of the functions $f_\nu$ have support in the
hypercube $\max_j(|p^0_j|,|p^1_j|)\le 4neb$, where
 $|P_c^2|\le 32 n^3e^2b^2$ and where $\omega(P_c^2/\Lambda^2)\equiv 1$
 with our choice of $\Lambda$. Therefore
$(\hat G_\Lambda,f_\nu)=(\hat G_{\Lambda_1},f_\nu)=(G,f_\nu)$ if
$b_1>b$, and hence $(\hat G_\Lambda,f)=(\hat G_{\Lambda_1},f)$.
 This consideration also shows that $\hat G$ is an extension of
    $G$ and is continuous in the topology of $S^0(\oU)$,
     which finishes the proof.

Theorem 7 shows that if we assume   weak form~\eqref{7.3} of the
spectral condition and  use the test function space $S^0$, then an
adequate modification of the microcausality axiom is the
requirement that  wedge~\eqref{7.4} be a strong carrier of the
matrix elements of the field commutators or anticommutators
 $\langle\Phi,\,[\phi_\iota(x),\phi_{\iota'}(x')]_\mp\Psi\rangle$
 for any $\Phi$ and $\Psi$  in a common dense invariant domain
$D$ of  fields in the Hilbert space of states.

We now show that the uniqueness theorem established in~\cite{2}
for the distributions supported by a properly convex cone allows a
direct generalization to the case  of a wedge-shaped support.

\medskip
 {\bf Theorem 8.} {\it Let $u\in \mathcal
D'(\oR^{d_1}\times\oR^{d_2})$ be a nontrivial distribution whose
support is contained in the wedge $V\times\oR^{d_2}$, where $V$ is
a properly convex cone. Then every strong carrier cone of its
Fourier transform $\tilde u\in
S^{\prime\,0}(\oR^{d_1}\times\oR^{d_2})$ has  the form
$\oR^{d_1}\times K_2$, where $K_2$ is a closed cone in
$\oR^{d_1}$.}

\medskip
{\bf Proof.} We assume that this is not the case and $\tilde u\in
S^{\prime\,0}(K_1,K_2)$, where $K_1\ne\oR^{d_1}$. Then for each
$g\in S^0(\oR^{d_2})$, the functional $\tilde u_g$ defined by
$(\tilde u_g,f)=(\tilde u, f\otimes g)$ is carried by the cone
$K_1$, as can be readily  seen from~\eqref{m}. The distribution on
$\oR^{d_1}$ whose Fourier transform is  $\tilde u_g$  has support
in $V$. Hence this distribution is trivial by Theorem 4
in~\cite{2}. Because $g$ is arbitrary and $\mathcal
D(\oR^{d_1})\otimes \mathcal D(\oR^{d_2})$ is dense in $\mathcal
D(\oR^{d_1}\times\oR^{d_2})$, we infer that $u\equiv 0$ and obtain
a contradiction.

Theorem 8 provides a way to extend some of the results
in~\cite{1}, \cite{2} on the spin-statistics relation and the CPT
symmetry to this version of  noncommutative field theory. Let
 \begin{equation}
 W_{\phi\phi^*}(x-x')=\langle\Psi_0,\,
 \phi(x),\phi^*(x')\Psi_0\rangle,\qquad
 W_{\phi^*\phi}(x-x')=\langle\Psi_0,\,
\phi^*(x),\phi(x')\Psi_0\rangle.
  \notag
\end{equation}
As a simplest example, we show that the wedge~\eqref{7.4} cannot
be a strong carrier of the anticommutator
$[\phi(x),\phi^*(x')]_+$, i.e., the anomalous commutation relation
is forbidden for the scalar field. Indeed,  the functional
$W_{\phi\phi^*}+W_{\phi^*\phi}$ is otherwise strongly carried by
the wedge $\bar \oV_c\times\oR^2$ by Theorem 7.
    From~\eqref{7.3}, it follows that the momentum-space support of this
    sum lies in the wedge $\bar \oV_{c+}\times\oR^2$.
  Therefore, this functional is  zero by Theorem~8. Averaging with a
   test function of the form $\bar f(x)f(x')$, we obtain $\|\phi^*(f)\Psi_0\|^2
   +\|\phi(f)\Psi_0\|^2=0$ and hence
    \begin{equation}
    \phi(f)\Psi_0=\phi^*(f)\Psi_0=0\qquad
   \text{for all}\quad f \in S^0(\oR^4).
     \notag
\end{equation}
As in the usual local theory~\cite{SW}, \cite{BLOT}, this implies
that the field $\phi$ vanishes. The arguments are analogous to
those in the proof of Theorem~13 in~\cite{1} and again use the
above generalization of the uniqueness theorem.

It is well known that the existence of CPT symmetry in the theory
of a scalar field is equivalent to the validity of the relations
   \begin{equation}
W(\xi_1,\dots,\xi_{n-1})= \breve W(\xi_{n-1},\dots,\xi_1),
  \label{7.12}
\end{equation}
  where $\breve W$ is defined by the same set of field operators as $W$ but
  taken in the inverse order. The functional
     \begin{equation}
W(\xi_1,\dots,\xi_{n-1})- \breve W(\xi_{n-1},\dots,\xi_1)
  \label{7.13}
\end{equation}
can be written as
\begin{equation}
 [W(\xi_1,\dots,\xi_{n-1})- \breve W(-\xi_{n-1},\dots,-\xi_1)]+
 [\breve W(-\xi_{n-1},\dots,-\xi_1)-\breve
   W(\xi_{n-1},\dots,\xi_1)].
    \label{7.14}
\end{equation}
According to Theorem~7, the expression in the second square
brackets in~\eqref{7.14} is strongly carried by the wedge
$\complement \oJ^c_{n-1}\times \oR^{2(n-1)}$. If the modified
causality condition is satisfied with the standard commutation
relation, then the functional in the first square brackets
in~\eqref{7.14} is also strongly carried by this wedge. In fact,
this functional expresses the difference between two vacuum
expectation values, one obtained from the other  by  permuting the
fields. This difference is representable as a sum of terms of the
form $\langle\Psi_0,\,\dots
[\phi^{(*)}(x_i),\phi^{(*)}(x_j)]_-\dots\Psi_0\rangle$, where the
dots stand for field operators. Such a term is strongly carried by
the wedge $\bar \oV_c\times\oR^{4n-2}=\{x \colon \,
(x_i-x_j)_c^2\geq 0\}$, whose image $\{\xi\colon\,
(\xi_i+\dots+\xi_{j-1})^2_c\geq 0\}$ in the relative coordinate
space is contained in $\complement \oJ^c_{n-1}\times\oR^{2(n-1)}$.
  It follows  from~\eqref{7.3} that the momentum-space support
  of  functional~\eqref{7.13} lies in the wedge  $\bar
\oV_{c+}^{n-1}\times\oR^{2(n-1)}$. Applying Theorem~8 again, we
conclude that the noncom\-mutative scalar field theory with the
functional domain $S^0(\oR^4)$, satisfying spectral
condition~\eqref{7.3} and the above-stated modification of
microcausality axiom,  has CPT as a symmetry. Moreover, the
presence of CPT symmetry is tantamount  to the condition that all
possible functionals in the first square brackets in~\eqref{7.14}
are strongly carried by the wedges $\complement \oJ^c_{n-1}\times
\oR^{2(n-1)}$, $n=2,3, \dots$, which is a generalization of weak
local commutativity.

\section{\large Conclusion}
The established  equivalence of  asymptotic commutativity to
certain regularity properties of the retarded  Green's functions
in momentum space  once again shows that the formulation of
nonlocal QFT  using  highly singular generalized functions is
quite rich in physical content because it has a self-consistent
interpretation in terms of particle scattering together with the
CPT symmetry and the standard spin-statistics relation. We note
that the distinction between carrier cones and strong carrier
cones is typical not only of the functional class $S^{\prime\,
0}$, but also of the generalized functions defined on any one of
the Gelfand-Shilov spaces $S^\beta$ and $S^\beta_\alpha$ with the
superscript $\beta\le 1$. This is particularly true in regard to
the Fourier hyperfunctions, which compose the dual space of
$S^1_1$ and provide the most general framework for constructing a
local field theory. We have commented on only one  attempt to
extend  the axiomatic approach to noncommutative field theories.
It will be instructive to consider the possibility of using
analytic test functions in other formulations, particularly in the
construction of Wightman functions  using the Moyal $*$-product
instead of the ordinary product of  fields and with the
implementation of  twisted Poincar\'e symmetry, as  proposed in
\cite{CK}, but this is beyond the scope of the present paper.

\medskip
\noindent {\bf Acknowledgments.} This paper was supported in part
by the the Russian Foundation for Basic Research (Grant
No.~05-01-01049) and the Program for Supporting Leading Scientific
Schools (Grant No.~LSS-1578.2003.2).

\end{document}